**Prebiotic Chemistry Insights for Dragonfly II: Thermodynamic Favorability of Nucleobases, Ribose, and Fatty Acids in Selk Crater on Titan**

Ishaan Madan[1*], Ben K.D. Pearce[1]

[1] Department of Earth, Atmospheric, and Planetary Sciences, Purdue University, West Lafayette, IN 47907, USA; madani@purdue.edu, pearce21@purdue.edu

*Corresponding Author

Prebiotic Chemistry in Selk Crater on Titan

**Abstract**
	Saturn's moon Titan is a prime destination for investigating prebiotic chemistry beyond Earth, particularly at impact crater sites where transient liquid water may have enabled aqueous reactions between organic molecules. Selk crater represents one such environment and is a primary target of NASA's Dragonfly mission. Here, we present a thermodynamic assessment of nucleobases, ribose, and fatty acids formed from simple atmospheric precursors (HCN and $C_2H_2$) within a Selk-sized aqueous melt pool across varying ammonia ($NH_3$) abundances. We find that ammonia acts as a chemical gatekeeper for molecular accessibility. In $NH_3$-free systems, accessibility is restricted to adenine and butanoic acid. Once ≥1% $NH_3$ is introduced, all investigated molecular classes become thermodynamically accessible. Distinct molecular classes have different $NH_3$ sensitivities: nucleobases, ribose, and $C_2$–$C_6$ fatty acids yield peaks at 1% $NH_3$, and $C_7$–$C_{12}$ fatty acids yield peaks at 2% $NH_3$. The modeled preference for pyrimidines vs. purines and monotonic decline of fatty acid abundance with chain length qualitatively mirror patterns observed in carbonaceous meteorites and returned asteroid samples. We show how molecular distributions and cross-class correlations may provide indirect constraints on Selk's past aqueous environment, help constrain past ammonia availability, and distinguish abiotic production from potential anomalies. By coupling thermodynamic predictions with an assessment of Dragonfly's mass spectrometer (DraMS) capabilities, we posit concrete, testable predictions for evaluating Selk's prebiotic potential in situ.

**1. Introduction**
	Saturn's largest moon, Titan, offers a uniquely rich natural laboratory for studying prebiotic chemistry beyond Earth. The Cassini–Huygens mission immensely elevated our understanding of Titan's chemical inventory, and a comprehensive synthesis of these mission results and subsequent studies is available in this recent book (Lopes et al. 2025). Titan's chemical factory begins in the upper atmosphere, where solar ultraviolet photons and energetic particles (originating from Saturn's magnetosphere) dissociate molecular nitrogen ($N_2$; ~95–98%) and methane ($CH_4$; ~2–5%) (Vuitton et al. 2025) to drive the synthesis of a wide range of organic species. Many of these products further polymerize into complex aerosol particles, giving rise to the thick global haze that shrouds the surface at optical wavelengths (e.g., Hörst 2017; Lora et al. 2025). Cassini observations, atmospheric models, and lab simulations suggest Titan's chemical inventory includes nitriles, hydrocarbons, and species relevant for prebiotic chemistry, which upon hydrolysis can yield biologically relevant compounds or their precursors, including amino acids and nucleobases (e.g., Khare et al. 1986; Raulin & Owen 2003; Niemann et al. 2005, 2010; Waite et al. 2007; Neish et al. 2009, 2010; Cable et al. 2012; Poch et al. 2012; Cleaves et al. 2014; Brassé et al. 2017; Pearce et al. 2024; Solomonidou et al. 2025; Vuitton et al. 2025). Over geological timescales, atmospheric chemical species (e.g., HCN and $C_2H_2$) adsorb onto organic haze particles that sediment onto the surface, producing an organic-rich layer blanketing Titan's water–ice crust (Coustenis 1997; Griffith et al. 2003; Janssen et al. 2016; Perrin et al. 2025). This atmospheric production and surface deposition is regarded as a dominant source of Titan's prebiotic chemical inventory. However, atmospheric haze chemistry alone does not define the upper bound of Titan's chemical potential.

While it provides a reservoir of complex organics, simpler molecules like hydrogen cyanide (HCN) and acetylene ($C_2H_2$) (along with other small nitriles and hydrocarbons; Krasnopolsky 2009) may deposit independently onto the surface where they can be mobilized and reprocessed under different environmental conditions (Neish et al. 2024). Titan's current cryogenic environment (surface temperatures of 89–94 K; Fulchignoni et al. 2005; Cottini et al. 2012; Jennings et al. 2019) precludes persistent surface liquid water; however, transient water-rich environments generated by impacts or putative cryovolcanic activity provide settings in which atmospheric organic deposition can interact with liquid water to drive further chemical evolution (Artemieva & Lunine 2003; Neish et al. 2018; Meyer-Dombard et al. 2025). Impact-generated melt pools may remain liquid for hundreds to thousands of years (Thompson & Sagan 1992; Artemieva & Lunine 2003; Hedgepeth et al. 2022) with recent modeling of Selk crater suggesting persistence for up to tens of thousands of years beneath an insulating ice shell (Wakita 脇田 et al. 2023; Kalousová et al. 2024). Selk crater is therefore a primary target of NASA's Dragonfly mission, scheduled to arrive at Titan in the mid-2030s (Barnes et al. 2021). Dragonfly's mass spectrometer, DraMS, will directly probe surface chemistry to assess the presence of biologically relevant molecules (Grubisic et al. 2021). Cassini remote-sensing observations indicate that Selk is compositionally heterogeneous rather than uniformly distributed with organics. Visual and Infrared Mapping Spectrometer (VIMS) analyses of equatorial dune craters show an organic-rich optical surface, and RADAR/radiometry and crater evolution studies imply that both shallow subsurface ice–organic mixing and the accessibility of exposed impact-melt materials vary laterally and vertically (Neish et al. 2018; Werynski et al. 2019; Solomonidou et al. 2020, 2024). At Selk crater, deeper portions of the melt may have persisted for several thousand years (Kalousová et al. 2024), but the upper layers most accessible to Dragonfly would have solidified on shorter timescales (0.5–600 yr, depending on melt depth) (Hedgepeth et al. 2022). In both cases, these environments permit aqueous processing of organics prior to refreezing, enabling chemical complexity beyond atmospheric pathways alone (e.g., Neish et al. 2018).

This work follows Madan and Pearce (2025), in which we used thermodynamic equilibrium models to explore amino acid synthesis within Selk-sized impact melt pools. Here, we extend that framework to encompass the remaining major molecular components of terrestrial life: nucleobases, ribose, and fatty acids. Together, these studies provide the first unified thermodynamic assessment of all four canonical classes of biological building blocks within a single Titan-relevant environment. Amino acids, nucleobases, and sugars are increasingly recognized as cosmochemically pervasive, having been identified in meteorites, asteroids, and primitive solar system materials (e.g., Engel & Nagy 1982; Cronin et al. 1995; Glavin et al. 1999, 2025; Glavin & Bada 2001; Martins & Sephton 2009; Cobb & Pudritz 2014; Cobb et al. 2015; Pearce & Pudritz 2016; Paschek et al. 2022; Naraoka et al. 2023; Oba et al. 2023; Potiszil et al. 2023a, 2023b; Sillerud 2024; Furukawa et al. 2025; Mojarro et al. 2025; Parker et al. 2025).

Prebiotic Chemistry in Selk Crater on Titan

On Titan, detection of such compounds would constrain how far prebiotic chemistry can progress in a hydrocarbon-rich, cryogenic world, and may begin to illuminate the boundary between simple organic synthesis and more advanced stages of chemical organization, which addresses core themes of the 2023-2032 Planetary Science and Astrobiology Decadal Survey (Committee on the Planetary Science and Astrobiology Decadal Survey et al. 2023). By beginning to establish abiotic baselines, we directly engage Priority Question 11 regarding the search for life. Our predictions for molecular favorability and distributions respond to Question 11.1a (*What is the Organic Molecule Inventory of Habitable Environments Throughout the Solar System, Including Complex Organic Molecules That Can Serve as Prebiotic Building Blocks of Life?*) and 11.1b (*What Is the Extent of Molecular Complexity…and Degree of Organization That Can Be Generated Abiotically…?*), demonstrating the chemical complexity expected from non-biological processes alone.

In this study, we evaluate the thermodynamic feasibility of the five canonical nucleobases (adenine, guanine, cytosine, uracil, thymine), two additional purine derivatives (xanthine and hypoxanthine), ribose (sugar in RNA), and saturated fatty acids from ethanoic through dodecanoic acid ($C_2$–$C_{12}$) within a Selk-sized melt pool, extending the methodology developed in Madan and Pearce (2025). These molecular classes are not expected to respond identically because they place different stoichiometric and energetic demands on the starting inventory. A unified thermodynamic assessment allows us to compare how shared environmental controls (e.g., ammonia abundance) shape the accessibility of distinct prebiotic molecule classes within the same melt pool model. We place the results in the context of the meteoritic record, provide testable predictions for Dragonfly, evaluate DraMS' suitability for detecting each molecular class, and outline how measured molecular distributions may be interpreted to distinguish expected abiotic chemistry from patterns that may warrant further chemical investigations into their origins.

**2. Methods**

In this companion study, we extend the thermodynamic modeling and Selk crater melt pool framework developed for amino acids (Madan & Pearce 2025) to nucleobases, ribose, and fatty acids. Unless otherwise noted, all physical assumptions and numerical methods are identical to those described in Section 2 of Madan and Pearce (2025). We detail our methods below.

*2.1. Thermodynamic Framework*

Thermodynamic modeling provides a principled way to estimate the maximum extent of chemical transformation by identifying the distribution of species that minimizes the total Gibbs free energy, $G$, of a system. At a constant temperature and pressure, $G$ represents the maximum non-mechanical (often chemical) work obtainable from a system after accounting for enthalpy of formation and entropic contributions. The equilibrium state therefore corresponds to the system configuration in which no further non-mechanical work can be extracted (minimum $G$), allowing thermodynamic models to predict product distributions independent of specific reaction pathways. Such models provide upper bound estimates of yields in chemical systems that may attain equilibrium and can be especially helpful to probe chemical evolution in systems that are not directly accessible (e.g., extraterrestrial environments) or challenging to study experimentally (e.g., toxic or unstable chemical systems). We compute chemical equilibrium using Cantera (v3.1.0; Goodwin et al. 2024), a Python-accessible thermodynamics and kinetics package. We employ the 'VCS' thermodynamic equilibrium solver (Missen & Smith 1998) which minimizes the total $G$ of the system at a fixed temperature, pressure, and bulk elemental abundances (i.e., a closed system) while ensuring the conservation of mass and charge. Conceptually, the solver asks: given a fixed number of atoms, what distribution of molecules yields the lowest total Gibbs free energy? It solves this problem by balancing the reactants and targeted product and computing the combination of species that results in the lowest overall $G$.

The thermodynamic framework requires Gibbs free energies of formation, $\Delta_f G^o(T)$, across a range of temperatures. For molecules with available literature data, we obtain $\Delta_f G^o(T)$ from the CHNOSZ database (Dick 2019) using the Python wrapper, pyCHNOSZ (Boyer 2024). For species lacking this data in CHNOSZ, namely xanthine and hypoxanthine, we estimate $\Delta_f G^o(T)$ using the Gibbs Free Energies Estimator described and validated in Section 2.2 of Madan and Pearce (2025). Quantum chemical calculations used in this estimator consist of geometry optimizations and frequency analyses performed at 298 K using density functional theory with the B3LYP hybrid functional (Lee et al. 1988; Becke 1993) and the 6-311++G(2df, 2p) basis set, as implemented in Gaussian 16 Revision B.01 (Frisch et al. 2017). Implicit solvation is modeled using the polarizable continuum model (PCM) with water as the solvent (Miertuš et al. 1981; Cammi & Tomasi 1995; Tomasi et al. 1999). Thermochemical quantities (electronic energies, zero-point energies, thermal corrections, entropic contributions) are extracted from the Gaussian output files and used in Ochterski's three-step method (Ochterski 2000) to compute $\Delta_f G^o$ at 298 K. As a continuous consistency check, we newly apply this method to adenine and find close agreement with the CHNOSZ value, with a deviation of 3.8% at 298 K. The validation and automation of this workflow were developed in Madan and Pearce (2025); the associated scripts are publicly available via Zenodo (Madan 2025). To reconstruct the temperature dependence of $\Delta_f G^o(T)$, the 298 K values are combined with temperature-dependent slopes derived from structurally analogous nucleobases: guanine for xanthine and adenine for hypoxanthine. This analog-based approach was benchmarked and validated in Madan and Pearce (2025). Table 1 summarizes the resulting $\Delta_f G^o(T)$ values from 0–100 ºC, with the full temperature range in Table A1 (Appendix A).

Table 1. Calculated Standard Gibbs Free Energies of Formation, $\Delta_f G^o$(J mol$^{-1}$) for Xanthine and Hypoxanthine.

| | $\Delta_f G^o$(J mol$^{-1}$) | |
|---|---|---|
| **T (ºC)** | **Xanthine** | **Hypoxanthine** |
| 0.01 | -104732 | 121455 |
| 10 | -108079 | 117898 |
| 20 | -111429 | 114338 |
| 30 | -114779 | 110778 |
| 40 | -118129 | 107218 |
| 50 | -121479 | 103658 |
| 60 | -124829 | 100098 |
| 70 | -128179 | 96538 |
| 80 | -131529 | 92978 |
| 90 | -134879 | 89418 |
| 100 | -138229 | 85858 |

Temperature primarily, in our models, governs the timescale over which equilibrium may be approached rather than reshaping the equilibrium identities over 0–100 ºC. Higher temperatures facilitate more rapid equilibration while lower temperatures may slow reaction rates; however, given sufficient time, the system converges toward the same minimum $G$ defined for that temperature. Over the explored range of 0–100 ºC, variations in $\Delta_f G^o(T)$ (~7–53 kJ mol$^{-1}$) are small relative to their absolute magnitudes and do not substantially alter equilibrium product identities in our parameterization used here.

To interface with Cantera, all $\Delta_f G^o(T)$ functions are encoded into NASA 9-coefficients (McBride et al. 2002), following the same fitting procedure used in Madan and Pearce (2025). The polynomial fitting script and example Cantera input file along with a model skeleton are archived on Zenodo (Madan 2025). The overarching steps of the methodology workflow are illustrated in Figure 1. We refer the interested reader to Madan and Pearce (2025) for additional conceptual details and validation of the thermodynamic framework and Gibbs Free Energies Estimator.

Figure 1. Overview of the thermodynamic modeling workflow used in this study and Madan and Pearce (2025).

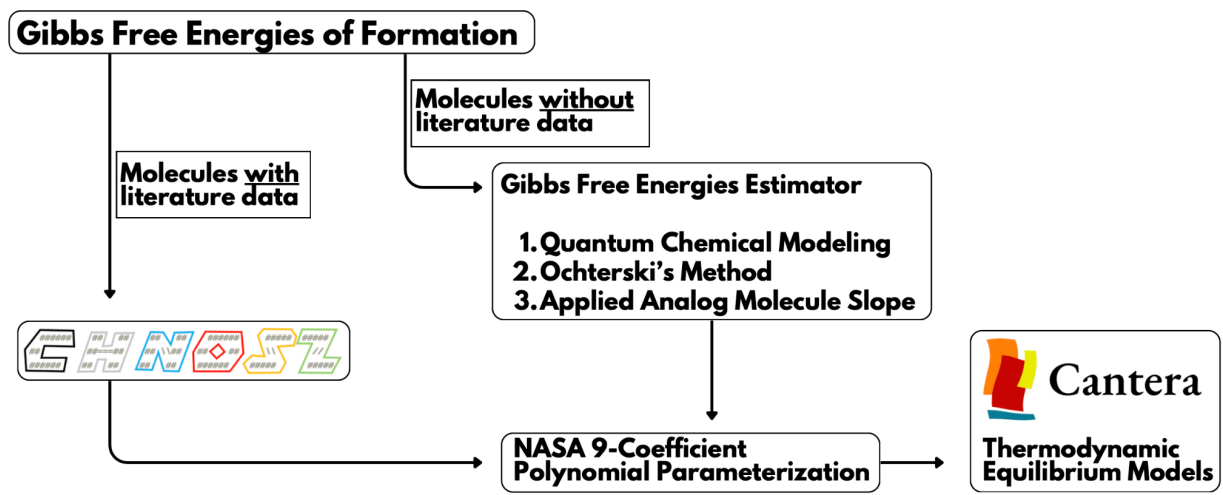

The reactant species include hydrogen cyanide (HCN), acetylene ($C_2H_2$), water ($H_2O$), and ammonia ($NH_3$). Our previous work (Madan & Pearce 2025) focused on amino acids as the modeled products and here we expand the target species list to include the five canonical nucleobases (adenine, guanine, cytosine, uracil, thymine), two additional purine derivatives (xanthine and hypoxanthine), ribose, and saturated fatty acids from ethanoic through dodecanoic acid ($C_2$–$C_{12}$). Each target product is modeled individually, utilizing a weak coupling approach. Weak coupling isolates thermodynamic favorability and mirrors experimental studies which target individual reaction systems without interference from an incompletely characterized chemical background. In contrast, modeling all target products in a single, simultaneous simulation (strong coupling) would rely on the assumption of a complete chemical landscape. Unless every byproduct and isomer is accounted for (a list potentially numbering in thousands), the Gibbs energy minimization would distribute mass inaccurately among the defined species.

*2.2. Selk Crater Melt Pool Model*

We adopt the Selk crater melt pool framework developed in Madan and Pearce (2025), which couples impact-generated melt volumes from hydrocode simulations (Wakita 脇田 et al. 2023) with atmospheric organic delivery estimates (Krasnopolsky 2009; Neish et al. 2024) to constrain plausible aqueous reaction environments on Titan. Simulations of a ~4 km diameter impactor into a methane–clathrate-rich ice shell suggest that a Selk-scale impact can generate on the order of 200–400 km³ of water–ice melt (Wakita 脇田 et al. 2023). The upper 10–50 m of the melt pool—corresponding to depths most accessible to Dragonfly—would require at least ~0.5–10 yrs to refreeze based on thermal evolution models that adopt a conservative 0 ºC melt with a HCN–water composition (Hedgepeth et al. 2022). Accounting for the presence of additional solutes and the possibility of elevated postimpact temperatures (>20 ºC), refreezing times for the upper layers may be extended. In contrast, deeper layers of the melt pool may remain liquid for several thousand to tens of thousands of years (Thompson & Sagan 1992; Artemieva & Lunine 2003), with estimates extending to ~25 kyr for sufficiently deep and insulated melts (Kalousová et al. 2024). These timescales motivate the central question of whether chemical equilibration is plausible within Selk crater's melt pond. As argued in Madan and Pearce (2025), experimental rate constants for aminonitrile hydrolysis experiments (Farnsworth et al. 2024) yield equilibration timescales ranging from ~1 to ~120 years (depending on temperature and concentration of ammonia; see Table 3 of Madan & Pearce 2025). These durations are orders of magnitude shorter than the freezing timescales of Selk-sized melt pools ($10^3$–$10^4$ years), suggesting that thermodynamic equilibrium is achievable, especially in the deeper, longer-lived melt layers. Partial equilibration may be possible in the shallower regions, especially if subjected to elevated postimpact temperatures (>20 ºC), supporting the plausibility of in situ thermodynamic equilibrium. Our models are time-independent, and the reported yields cannot be constrained more precisely in time and should be interpreted as equilibrium accessibility and upper-bound reference states within the broad melt-lifetime ranges discussed here. When comparing Selk (~90 km diameter) to Titan's largest crater Menrva (~425 km diameter), formation and evolution modeling of the latter show complete and partial melt extending ~65 km depth (Crósta et al. 2021). In most of their simulations, the crater breaches the ice shell so that subsurface ocean water is brought to depths as shallow as ~10 km. Relative to Menrva, Selk likely represents a smaller and less extreme impact-habitability scenario, and the comparison serves as a broader environmental context for the melt conditions modeled in this work.

      Cassini observations provide geological context for the simplified Selk crater melt pool model adopted in this work. VIMS constrains Titan's optical surface layer, i.e., the top few microns, and RADAR emissivity constrains the shallow subsurface from a few tens of centimeters to roughly a meter depth (Werynski et al. 2019; Solomonidou et al. 2020, 2024). VIMS-based compositional analyses of equatorial dune craters, including Selk, are dominated at the optical surface by organic materials and show no evidence for $H_2O$, $NH_3$, or $CO_2$ ice, consistent with Selk's dune setting (Solomonidou et al. 2020). RADAR/radiometry results and crater-evolution studies indicate a more heterogeneous shallow subsurface with the crater and ejecta in dune settings interpreted to retain a greater water-ice contribution immediately after the impact, and crater floors becoming more organically enriched through subsequent aeolian infilling and degradation (Werynski et al. 2019; Solomonidou et al. 2020). The most accessible impact-melt-bearing materials at such Titan craters are likely to be localized rather than laterally uniform, with fresher crater floor exposures being the most promising target for Dragonfly (Neish et al. 2018). These observations indicate lateral and vertical variability in organic cover, shallow subsurface ice–organic mixing, and the accessibility of impact-melt materials at Selk. Our thermochemical model assumes a well-mixed, homogeneous aqueous melt; we return to this assumption in Section 2.3. These observations are included to bound the realism of the initial organic inventory and melt-mixing assumptions rather than being represented explicitly in the model.

      Initial reactant concentrations are inherited directly from Madan and Pearce (2025), where atmospheric organic deposition fluxes (Neish et al. 2024), rooted in photochemical models (Krasnopolsky 2009), were combined with Selk's catchment area and steady deposition over geological timescales based on age estimates of Titan's surface and Selk crater (Neish & Lorenz

2012; Neish et al. 2015, 2016; Werynski et al. 2019; Hedgepeth et al. 2020, 2022; Solomonidou et al. 2020; Lorenz et al. 2021; Wakita et al. 2024). Deposited inventories of HCN and $C_2H_2$ are normalized to the melt pool water concentration implied by the adopted 200 km$^3$ melt volume. A 10% survival factor is applied to both species to account for any destruction or incomplete incorporation into the melt pool. This choice is consistent with our previous study and with recent Titan impact studies that estimated total amino acid survival fractions of ~12–33% under average impact conditions (Pearce et al. 2024). To further account for uncertainty regarding organic survival and mixing efficiency, we perform sensitivity analyses using survival factors of 1% and 30% (Appendix B, Figure B1, Tables B1 and B2), which bracket conservative and optimistic scenarios. Ammonia likely played a key role in Titan's early evolution, interior, and subsurface ocean chemistry (Lunine et al. 2025), but its abundance at Titan's surface is poorly constrained observationally, and estimates for subsurface liquid reservoirs remain model-dependent. Surface studies from the Cassini-era have not confidently identified $NH_3$ ice at Titan's surface and compositional analyses of impact craters find no evidence for $NH_3$ ice at Selk or related crater terrains (Solomonidou et al. 2020). This is in part because Titan's thick haze and limited near-infrared atmospheric windows make surface compositional retrievals difficult (Nasralla et al. 2025). Interior models span a broad range: early estimates based on cosmic abundances placed the ammonia content of the subsurface ocean at ~0.16 mole fraction or roughly ~15 wt.% (Lunine & Stevenson 1987; Cynn et al. 1989), with geochemical models adopting similar values (Engel et al. 1994). Subsequent Cassini-era modeling explored lower concentrations, considering up to 14 wt.% $NH_3$ beneath a water-ice shell (Tobie et al. 2005) and a maximum of ~5 wt.% $NH_3$ in later ocean-composition models (Leitner & Lunine 2019). Experimental studies on Titan-relevant aqueous chemistry have explored overlapping $NH_3$ concentration ranges: 3–25 wt.% (Poch et al. 2012), 5–15 wt.% (Farnsworth et al. 2024), and 20.5 wt.% solutions (Nasralla et al. 2025). Recent geophysical work argues against a present-day global subsurface ocean, although it permits a slushy high-pressure ice layer and possible localized melt pockets, further complicating simple estimates of $NH_3$ availability to impact-generated melts (Petricca et al. 2025). If ammonia is present in Titan's interior, it may be mobilized into an impact melt via direct release from the interior or disruption of $NH_3$-bearing ices and clathrates during impact (Lunine & Stevenson 1987; Choukroun & Sotin 2012; Kalousová & Sotin 2020). However, its effective concentration in an impact melt depends on uncertain partitioning, excavation depth, dilution within the melt volume, and possible enrichment during refreezing. These studies suggest that percent-level $NH_3$ in an impact-generated aqueous system is geologically plausible but not directly constrained for Selk crater or related terrains, where dilution and uncertain partitioning may substantially reduce effective concentrations relative to source-region values. We therefore treat $NH_3$ as a poorly constrained yet geologically plausible variable and explore 0–10% $NH_3$ (relative to water), with 0% representing complete absence or loss of ammonia. All reactant concentrations used in the fiducial models are summarized in Table 2.

Table 2. Initial concentrations of the reactants used in our fiducial thermochemical models.

| Molecule | Concentration (mol X/mol $H_2O$) |
|---|---|
| $H_2O$ | 1.00 |
| $C_2H_2$ | 0.042 |
| HCN | 0.020 |
| $NH_3$ | 0, 0.01, 0.02, 0.03, 0.04, 0.05, 0.10 |

*2.3. Assumptions, Limitations, and Caveats*

As in the previous work with amino acids (see Section 3.4 of Madan & Pearce 2025), several simplifying assumptions are inherent to this analysis, which is designed to explore the thermodynamic feasibility of nucleobases, ribose, and fatty acids rather than to predict explicit reaction pathways or absolute abundances. HCN and C2H2 are the focus of the initial organic inventory for two reasons: (i) we seek the minimum viable organic inventory that may produce biomolecules, and (ii) these are among the most abundant and tractable species constrained by current estimates of Titan's surface organic inventory (Krasnopolsky 2009; Neish et al. 2024). Complex organic haze fractions ($C_xH_y$, $C_xH_yN$), which constitute a major organic reservoir on Titan, are excluded from the starting inventory due to poorly defined molecular identities and lack of thermodynamic data; this treatment is consistent with our previous amino acid study (Madan & Pearce 2025). Given that Titan's atmospheric haze delivers a broader and more complex organic inventory than the minimal starting inventory explored here, the overall magnitude and diversity of organic synthesis within the melt pool could plausibly exceed the bounds established by our fiducial models.

We assume an ideal aqueous solution baseline with unit activity coefficients, thereby neglecting non-ideal effects arising from ionic strength, ion pairing, or specific solute-solute interactions. The melt pool is treated as a single, homogeneous aqueous phase, without explicit treatment of volatile loss, phase separation, or spatial heterogeneity, despite observational evidence that Selk is laterally and vertically variable at depths accessible to Dragonfly (Neish et al. 2018; Werynski et al. 2019; Solomonidou et al. 2020, 2024). These assumptions are expected to affect the quantitative results more than the qualitative trends. For example, non-ideal solution chemistry may shift the precise threshold concentrations at which some products become favorable and melt pool heterogeneity could produce local enrichment or depletion relative to the bulk equilibrium composition. Acid–base speciation is not tracked explicitly; while ammonia controls alkalinity, pH is not independently solved or varied. We treat freezing-point depression qualitatively and neglect explicit phase equilibria associated with freezing, including ice–brine partitioning and solute exclusion during solidification. We assume freshwater melt pools, without added salts or minerals, but impact melts on Titan may incorporate crustal materials which could depress freezing temperatures, buffer pH, promote metal-organic complexation, and/or catalyze reactions. Incorporating such effects would likely modify both melt lifetimes and chemical speciation, representing an important avenue for future work. In a real Selk melt environment, such processes could generate vertical and lateral chemical gradients during progressive refreezing, making shallower materials, which are most accessible to Dragonfly, chemically distinct from deeper, longer-lived melt zones (e.g., Hedgepeth et al. 2022). Additional uncertainties arise from the initial reactant inventories, survival fractions, mixing efficiencies, and the availability of ammonia. The sensitivity tests partially bracket uncertainty in organic survival and melt incorporation, but they do not resolve non-ideal solution chemistry or freeze-out-driven redistribution.

Knowing thermodynamics alone does not guarantee that a reaction will occur. Kinetics govern specific reaction pathways, rates of reactions, and energy barriers that must be overcome. However, constructing kinetic models for Titan's surface chemistry is currently hindered by significant data gaps. Specifically, we lack precise constraints on the starting organic chemical inventory, the specific reaction networks for complex synthesis, and the requisite rate constants for these chemical transformations in Titan-relevant environments (e.g., extremely low temperatures). Elucidating a single formation pathway to the level of detail required for kinetic modeling is resource intensive. Recent experiments, for example, focused specifically on the hydrolysis of aminonitriles to form amino acids (Farnsworth et al. 2024). By the same token, a computational study had to resolve complex non-equilibrium interplays between multiple intermediates and calculate rates for dozens of individual elementary steps just to define the mechanism for adenine synthesis from HCN (Cappelletti & Rahm 2026). Scaling such resource-intensive approaches—whether long-duration experiments or rigorous computational modeling—to the thousands of potential species in a complex prebiotic soup would require prohibitive timescales and resources. We therefore use thermodynamic modeling as a first-pass filter to identify what chemistry is energetically permitted under Selk-relevant conditions and to highlight broad constraints that remain informative irrespective of the precise pathway. For example, our results highlight the thermodynamic importance of ammonia for the formation of most nucleobases and ribose, a constraint that remains valid regardless of the specific kinetic pathway taken. Ideally, the thermodynamic predictions should be validated by targeted laboratory experiments; we encourage this as an open avenue for future work. Since we also don't consider photolysis, radiolysis, mineral catalysis, competing reactions for polymerization, adsorption, and oxidative sinks in our models, future work can direct focus here as well. For instance, ongoing experiments are investigating the impact of phosphate salts and gamma-ray irradiation (proxy for Galactic Cosmic Ray deposition) on aqueous organic chemistry relevant to Selk crater (Masson et al. 2025). Their preliminary results suggest that factors such as mineral catalysis and ionizing radiation could lower kinetic barriers, potentially accelerating reactions that are thermodynamically favorable but kinetically sluggish. By combining thermodynamic assessments with experimental kinetic data—as demonstrated in our comparison of glycine and alanine yields (Madan & Pearce 2025) with laboratory reaction rates (Farnsworth et al. 2024)—we can build upon and strengthen the predictive framework for what Dragonfly might detect on Titan's surface.

This study falls within the growing field of computational astrobiology, where modeling approaches are needed and utilized to bridge substantial observational and experimental gaps (e.g., poorly constrained surface inventories, limited data for complex organics, and inaccessible planetary environments). In such contexts, explicitly stating and bounding assumptions is also crucial so that results are interpreted appropriately and not overgeneralized. As conceptually emphasized in Madan et al. (2025), such computational work should be considered tools to map what chemistry is thermodynamically permitted under plausible extraterrestrial conditions. Despite the limitations, the present results provide a physically consistent basis for comparing the thermodynamic feasibility of nucleobases, ribose, and fatty acids in a Selk-sized melt pool. When combined with lab experiments, kinetic constraints, and future in situ measurements—such as those anticipated from the Dragonfly mission—these results are a step toward bounding the potential for complex organic synthesis on Titan's surface.

## 3. Results

Percent yields for the nucleobases (Fig. 2A), ribose sugar (Fig. 2B), and $C_2$–$C_{12}$ fatty acids (Fig. 2C) are shown in Figure 2 and reported in Table 3. HCN is the limiting reagent for nucleobases and ribose, and $C_2H_2$ is the limiting reagent for fatty acid synthesis in the fiducial models. At 0% $NH_3$, only adenine and butanoic acid are thermodynamically accessible. All other studied molecules become accessible once 1% $NH_3$ is introduced, with yields peaking at that mark and tapering thereafter. The exceptions to this trend are $C_7$+ fatty acids, whose yields peak at 2% $NH_3$ and taper thereafter (Table 3).

Figure 2. Heat maps of percent yields (applicable to aqueous environments at thermodynamic equilibrium spanning 0–100 ºC) relative to the starting limiting reagent for (A) nucleobases, (B) ribose, and (C) $C_2$–$C_{12}$ fatty acids in our fiducial models. Columns correspond to varying $NH_3$ concentrations (relative to total water content). Rows list the individual molecules in each class. Cell shading follows a common viridis color scale from 0% (purple) to >70% (yellow) yield.

Prebiotic Chemistry in Selk Crater on Titan

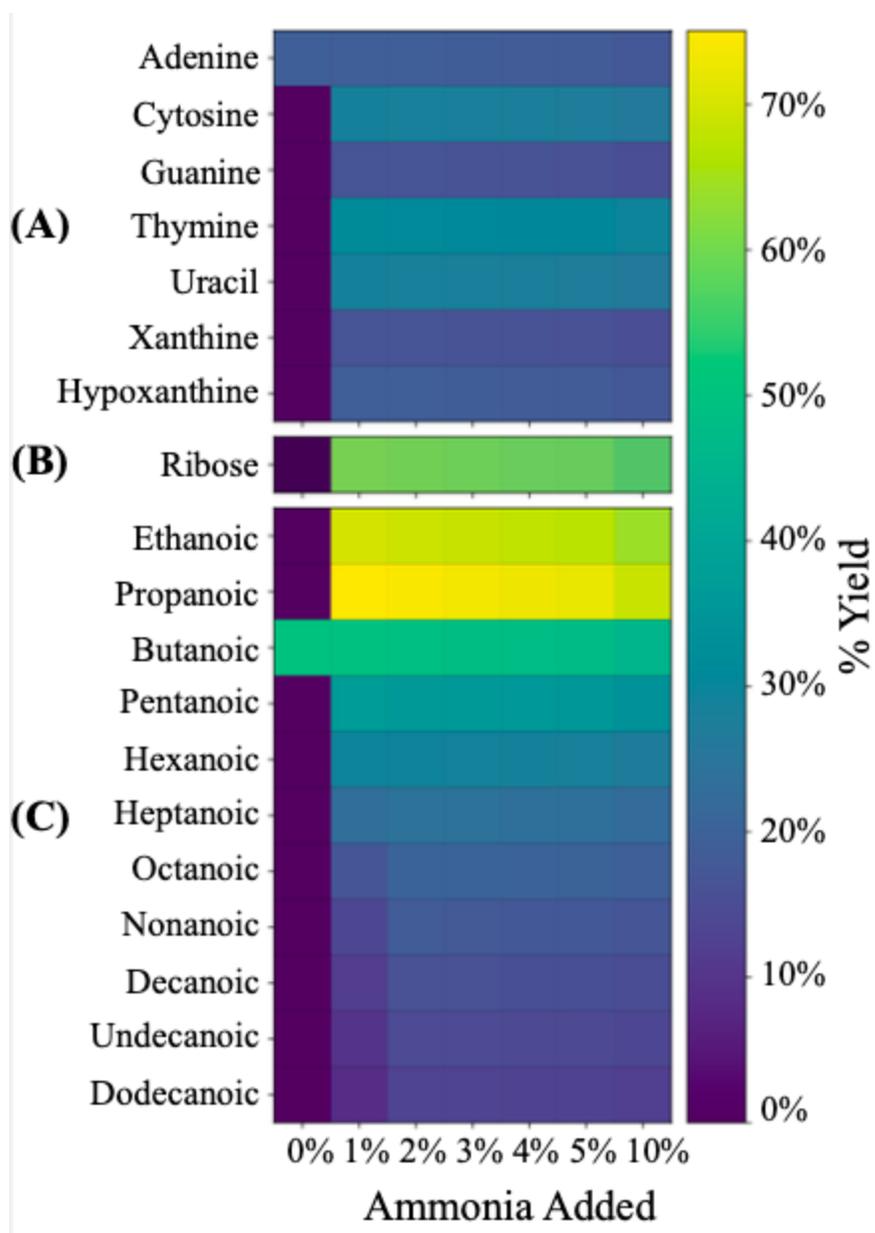

Table 3. Percent yields of nucleobases, ribose, and $C_2$–$C_{12}$ fatty acids, expressed relative to the limiting reagent (HCN for nucleobases and ribose; $C_2H_2$ for fatty acids).

| Molecule Class | Molecule | Yield | | | | | | |
|---|---|---|---|---|---|---|---|---|
| | | 0% $NH_3$ | 1% $NH_3$ | 2% $NH_3$ | 3% $NH_3$ | 4% $NH_3$ | 5% $NH_3$ | 10% $NH_3$ |
| Nucleobase | Adenine | 19.1 | 18.9 | 18.8 | 18.6 | 18.4 | 18.2 | 17.5 |
| | Cytosine | 0.0 | 28.5 | 28.2 | 28.0 | 27.7 | 27.5 | 26.3 |
| | Guanine | 0.0 | 16.7 | 16.6 | 16.4 | 16.3 | 16.1 | 15.4 |

| | | | | | | | |
|---|---|---|---|---|---|---|---|
| | Thymine | 0.0 | 31.9 | 31.6 | 31.3 | 31.0 | 30.7 | 29.4 |
| | Uracil | 0.0 | 28.5 | 28.2 | 28.0 | 27.7 | 27.5 | 26.3 |
| | Xanthine | 0.0 | 16.7 | 16.6 | 16.4 | 16.3 | 16.1 | 15.4 |
| | Hypoxanthine | 0.0 | 18.9 | 18.8 | 18.6 | 18.4 | 18.2 | 17.5 |
| **Sugar** | Ribose | 0.0 | 59.8 | 59.2 | 58.6 | 58.0 | 57.5 | 54.8 |
| | Ethanoic | 0.0 | 69.9 | 69.2 | 68.5 | 67.9 | 67.3 | 64.2 |
| | Propanoic | 0.0 | 75.1 | 74.4 | 73.6 | 72.9 | 72.2 | 68.9 |
| | Butanoic | 50.1 | 49.6 | 49.1 | 48.6 | 48.1 | 47.7 | 45.5 |
| | Pentanoic | 0.0 | 37.0 | 36.6 | 36.3 | 35.9 | 35.6 | 34.0 |
| | Hexanoic | 0.0 | 29.5 | 29.2 | 28.9 | 28.6 | 28.4 | 27.1 |
| **Fatty Acid** | Heptanoic | 0.0 | 23.3 | 24.3 | 24.1 | 23.8 | 23.6 | 22.5 |
| | Octanoic | 0.0 | 17.3 | 20.8 | 20.6 | 20.4 | 20.2 | 19.3 |
| | Nonanoic | 0.0 | 13.8 | 18.2 | 18.0 | 17.8 | 17.7 | 16.9 |
| | Decanoic | 0.0 | 11.5 | 16.1 | 16.0 | 15.8 | 15.7 | 15.0 |
| | Undecanoic | 0.0 | 9.8 | 14.5 | 14.4 | 14.2 | 14.1 | 13.5 |
| | Dodecanoic | 0.0 | 8.6 | 13.2 | 13.1 | 12.9 | 12.8 | 12.2 |

*3.1. Sensitivity Analysis*

Across the explored survival range (1%, 10%, 30%) of the initial organic inventory, uncertainties primarily affect the yield distributions: higher organic survival slightly decreases nucleobase yields while marginally increasing ribose production. The qualitative trends of the fiducial model (10% survival) are maintained across all species, with the notable exception of longer-chain fatty acids (especially $C_7+$) in the 30% survival scenario. In this case, yields are suppressed, shifting the optimal production window to higher ammonia abundances (~2–5% $NH_3$, increasing with chain length) (Appendix B, Figure B1, Tables B1 and B2).

**4. Discussion**

The results suggest that the formation of amino acids (Madan & Pearce 2025), nucleobases, ribose, and fatty acids in the models simulating Selk crater's melt pool environment is governed by three coupled factors: (i) elemental stoichiometry, (ii) initial organic inventory, and (iii) Gibbs energetics. Although liquid water is abundant in the model, hydrogen availability emerges as a thermodynamic constraint because hydrogen is largely sequestered in a stable, oxidized reservoir, making its redistribution into reduced organic products energetically costly. The introduction of ammonia therefore plays a central role in efficiently supplying hydrogen, enabling the system to satisfy atomic balance while minimizing free energy and shaping which molecular classes can be assembled. In parallel, the reduced carbon feedstocks delivered from the atmosphere, namely HCN and $C_2H_2$ in our models, define accessible reaction backbones and bias the system toward specific molecular families. Thermodynamic favorability then determines how these stoichiometric and compositional constraints are resolved, selecting products that minimize the total Gibbs free energy. Together, these factors underlie and define the behavior of the equilibrium solver and provide necessary explanations for interpreting the patterns of prebiotic chemistry predicted for Selk crater.

*4.1. Nucleobases: Adenine, Cytosine, Guanine, Thymine, Uracil, Xanthine, Hypoxanthine*

Adenine is the only nucleobase produced in the absence of ammonia. This is consistent with the HCN oligomerization pathway in which HCN polymerizes to form adenine (Oro & Kimball 1961; Ferris et al. 1978; LaRowe & Regnier 2008; Cappelletti & Rahm 2026) because the route is stoichiometrically balanced in C, H, and N. In experimental work of Oro and Kimball (1961), the formation of adenine from HCN is facilitated with ammonia, whereas Ferris et al. (1978) demonstrated HCN and $H_2O$ as the reactants for adenine and modeling efforts of LaRowe and Regnier (2008) further included a direct oligomerization pathway of aqueous HCN into aqueous adenine, providing a close equilibrium-based comparison to our approach. More recently, quantum chemical and microkinetic analyses have shown that adenine represents the thermodynamic sink of HCN self-reaction chemistry, occupying the global minimum on the Gibbs energy landscape of C, H, N compounds with 1:1:1 stoichiometry (Cappelletti & Rahm 2026). Aligning with our model, HCN alone supplies the full complement of hydrogen, carbon, and nitrogen atoms required for adenine, allowing for adenine to form. As a result, adenine yields in our models are largely insensitive to the ammonia abundance and reflect the self-contained stoichiometry of this reaction network.

In contrast, the remaining nucleobases (cytosine, uracil, thymine, guanine, xanthine, and hypoxanthine) are only produced once ammonia is introduced into the system. Across all $NH_3$-bearing models, the nucleobase distribution exhibits a clear preference for pyrimidines over purines, following the ranking: thymine > (cytosine, uracil) > (adenine, hypoxanthine) > (guanine, xanthine). Crucially, Koga et al. (2025) suggest a direct correlation between ammonia abundance and nucleobase distributions. The contrast in nucleobase ranking depending on ammonia availability is remarkably consistent with recent sample return analyses from carbonaceous asteroids. Samples from asteroid Bennu contain water-extractable ammonia abundances 75 times higher than those from asteroid Ryugu (Glavin et al. 2025). Ammonia-poor Ryugu favors purines, with a purine-to-pyrimidine ratio of ~1.1-1.2. Bennu, with higher ammonia, shifts to favoring pyrimidines with a ratio of ~0.55. Orgueil, another carbonaceous chondrite with the highest ammonia content (Laize-Générat et al. 2024; Glavin et al. 2025) exhibits the most extreme shift, with a ratio of ~0.099 (Koga et al. 2025). This observational gradient—where increasing ammonia concentrations progressively suppress purine dominance in favor of pyrimidines—validates our model's prediction that high ammonia availability thermodynamically drives the system toward pyrimidine synthesis.

Furthermore, while some arguments highlight the kinetic instability of cytosine and its deamination to uracil (e.g., Pearce & Pudritz 2016), high-sensitivity detections by Oba et al. (2022) have confirmed the presence of all five canonical nucleobases (including cytosine and uracil) in Murchison, Murray, and Tagish Lake meteorites. This suggests that prebiotic environments can indeed support the formation and preservation of pyrimidines, potentially through steady-state production or protective mineral associations, consistent with the thermodynamic favorability we observe here.

The full enumeration of the individual reaction pathways is beyond the scope of this work. However, limited further probing demonstrates that it's possible for formaldehyde ($CH_2O$) to play a role in forming cytosine, uracil, thymine, guanine, xanthine, and hypoxanthine in the presence of ammonia. Additional targeted equilibrium calculations, in which we explicitly include $CH_2O$ as a candidate species, show that $CH_2O$ is not thermodynamically favored in the absence of ammonia but becomes accessible once ammonia is introduced. This provides a potential mechanistic explanation for the suppression of nucleobases (except adenine) in $NH_3$-free systems: without ammonia, the system cannot produce the formaldehyde-like precursors necessary for pyrimidine synthesis (as detailed in Table 1 of Pearce & Pudritz 2016).

*4.2. Sugar: Ribose*

Ribose is only produced in our models once ammonia is present. This behavior is primarily driven by stoichiometric hydrogen limitations rather than carbon availability. By definition, sugars obey C:H:O ratios of 1:2:1 (e.g., ribose: $C_5H_{10}O_5$), which demands two hydrogens per carbon. In our initial organic inventory, both HCN and $C_2H_2$ only carry one hydrogen per carbon. Under $NH_3$-free conditions, these reactants are hydrogen-poor relative to the requirements of sugar formation. As the equilibrium solver must satisfy atomic balance while minimizing Gibbs free energy, the formation of highly hydrogenated species such as ribose is penalized when hydrogen is limiting.

The $NH_3$ dependence lines up with our isolated testing behavior for $CH_2O$: when $NH_3$ is absent, $CH_2O$ is not produced, removing a potential intermediate for sugar chemistry. This is particularly important to note here since the best-known abiotic route to pentoses (five-carbon sugars) is the formose (Butlerov) reaction, which begins with $CH_2O$ and builds larger sugars through aldol/retro-aldol cycling (Butlerow 1861). The suppression of $CH_2O$ under $NH_3$-free conditions in our testing therefore provides a chemically plausible explanation for the absence of ribose in those cases, similar to the case of nucleobases, with the exception of adenine as explained.

Our thermodynamic threshold is again validated by recent analysis of returned asteroid samples which exhibit a dichotomy that mirrors our model results. In the analysis of samples from asteroid Ryugu (ammonia-poor), Meinert et al. (2025) detected the aldopentose sugars arabinose and xylose; but notably, no ribose. This non-detection must be viewed with caution because their recovery tests indicated an extremely poor extraction efficiency of ribose (~4.5%), meaning that even if ribose was present, it would likely fall below detection limits in their small Ryugu sample. Therefore, non-detection may not mean absence definitively. Whereas, in the Orgueil sample (ammonia-rich), they detect ribose (at near-racemic levels suggesting it is not contaminant) (Meinert et al. 2025). Similarly, Furukawa et al. (2025) successfully identified ribose (along with arabinose and xylose, similar to Ryugu) in samples from asteroid Bennu. Given that Bennu and Orgueil both contain higher ammonia abundances than Ryugu (as discussed in Section 4.1), this observational split supports our model result that ribose is not thermodynamically prohibited under $NH_3$-rich conditions.

Our findings are also consistent with the broader thermochemical modeling of aqueous planetesimals. While the formose reaction is kinetically complex and prone to tar formation, ribose is thermodynamically favorable in aqueous conditions if the precursors are available (Paschek et al. 2022). While their reactant inventory and environmental assumptions differ substantially from the Selk crater melt pool model explored here, the qualitative result of ribose favorability is consistent. Our results align with this broader thermochemical expectation, with ammonia acting as the enabling factor that permits sufficient hydrogen balance and consequently, satisfies the elemental ratios and energetics required for ribose formation in our models.

*4.3. Fatty Acids: Ethanoic ($C_2$) through Dodecanoic ($C_{12}$)*

Butanoic acid ($C_4$) is the sole fatty acid product in an $NH_3$-free scenario, with other chain lengths—including the simpler ethanoic ($C_2$) and propanoic ($C_3$) acids—entirely suppressed. The formation of butanoic acid in the absence of ammonia can be explained by a stoichiometrically favorable coupling of acetylene with water (e.g., $2\ C_2H_2 + 2\ H_2O \rightarrow C_4H_8O_2$, Butanoic Acid) where the minimization solver can perfectly balance carbon, hydrogen, and oxygen without requiring additional hydrogen. In contrast, the remaining fatty acids strictly require ammonia presence, despite containing no nitrogen in their structure. This dependence stems from the hydrogen budget of the system. Fatty acids are saturated hydrocarbons (no double bonded carbons) with a carboxyl group (–COOH), generally requiring $2n + 1$ hydrogens for every carbon ($C_nH_{2n+1}COOH$, where *n* is the number of carbons). Our starting organic inventory, in an $NH_3$-free scenario, is hydrogen-poor: both HCN and $C_2H_2$ possess a 1:1 H:C ratio. As the carbon chain lengthens, the system faces a steep hydrogen deficit. Consequently, the full suite of fatty acids is only thermodynamically accessible once ammonia is introduced. Ammonia acts as a hydrogen reservoir, offering three hydrogens for each molecule and increasing the system's overall H:C ratio.

Similar to our model, in the meteoritic record, carbonaceous chondrites exhibit a distribution of fatty acids that generally decreases with increasing carbon number—a sign of abiotic synthesis (Sephton 2002 and references therein). More recently, Lai et al. (2019) compiled abundances from various meteorite studies, confirming that when fatty acids are preserved, they appear as a homologous series. Here, homologous series refers to a set of similar, successive fatty acids that differ from one another by a single methylene (–CH$_2$–) unit. While meteorites contain diverse isomers, we restrict our comparison here to the straight-chain monocarboxylic acids (SCMA's) investigated in our models. Our model results under NH$_3$-bearing conditions reproduce some features of the SCMA distribution in the Murchison meteorite: (1) propanoic acid being the most abundant (Sephton 2002) and (2) monotonic decrease in yields as the carbon chain lengthens from C$_4$–C$_{12}$. The latter trend mirrors the aforementioned general decrease in abundance with increasing chain lengths (Sephton 2002; Lai et al. 2019). A notable nuance in our results is the C$_2$ vs. C$_3$ relationship. While standard synthesis (e.g., Fischer-Tropsch) typically predicts a strict monotonic decrease (C$_2$ > C$_3$ > C$_4$), our thermodynamic model favors propanoic acid over ethanoic acid (C$_3$ > C$_2$). This inversion likely reflects the specific stoichiometry constraints in the model imposed by our starting inventory (HCN, C$_2$H$_2$) which seems to favor propanoic acid. However, the agreement with the relatively high propanoic acid concentration reported in Murchison suggests that such deviations from strict monotonicity are physically plausible and may reflect specific parent body precursor inventories.

The meteorite and parent-body comparisons presented in Sections 4.1–4.3 remain largely qualitative since parent-body aqueous alteration differs substantially from transient impact-melt processes on Titan (e.g., reactant inventory, duration of aqueous phase).

*4.4. Implications for Dragonfly and Other Missions*

Dragonfly's mass spectrometer, DraMS, is designed to investigate the chemical complexity of Titan's surface using two primary modes: Laser Desorption Mass Spectrometry (LDMS) for broad molecular surveying of refractory organics, and Gas Chromatography-Mass Spectrometry (GC-MS) for the separation and identification of functionalized molecules (i.e., compounds containing groups such as –COOH, –NH$_2$, or –OH) (Grubisic et al. 2021). Because DraMS is optimized to analyze shallow solid surface materials, ammonia may be difficult to constrain from any single measurement. Its present-day abundance in a sampled solid could have been modified by dilution, volatile loss, refreezing, or redistribution after the melt pool formed. Our thermodynamic results therefore suggest that the distributions of nucleobases and fatty acids, together with any detection of ribose, may provide more useful indirect constraints on the environmental conditions of the melt pool's aqueous phase, including past ammonia availability. We remind the reader that our model is time-independent, and these predictions should be interpreted as thermodynamic baselines and upper-bound reference states (with respect to the starting organic inventory explored) rather than exact abundances expected in any single Dragonfly sample. Shallow materials accessible to Dragonfly may preserve only partial equilibration and may differ from the bulk composition of deeper, longer-lived melt zones. Below, we propose five specific, testable predictions for Dragonfly along with their respective detection strategies.

First, given our nucleobases results, the relative intensity of purine versus pyrimidine signals in LDMS spectra can serve as a proxy for ammonia availability. Our models indicate adenine formation without ammonia (via HCN polymerization), while the presence of ammonia can shift the system to favor pyrimidines (thymine, cytosine, uracil). The DraMS development team has explicitly demonstrated the identification of adenine in LDMS spectra (Stern et al. 2023). For further definitive structural identification, GC-MS can also be employed using one of their three pre-treatments planned to volatilize polar or refractory compounds: thermochemolysis with tetramethylammonium hydroxide (TMAH) (He et al. 2019; Boulesteix et al. 2024a; Freissinet et al. 2024). The other two pre-treatments for GC-MS are pyrolysis (at 600 ºC for a few seconds) and derivatization using DMF-DMA (N, N-DiMethylFormamide DimMethylAcetal), with the latter more suited for analysis of amines, carboxylic acids, and chiral analysis (Boulesteix et al. 2024b). The effectiveness of thermochemolysis with TMAH has been demonstrated for Sample Analysis at Mars (SAM) instrument of the Mars Curiosity Rover and the Mars Organic Molecule Analyzer (MOMA) instrument of the ExoMars Rosalind Franklin Rover (He et al. 2019), which serve as heritage instruments for DraMS (Grubisic et al. 2021). Consequently, we expect that an LDMS or TMAH pre-treated GC-MS spectra can be used for detecting nucleobases, where the purine-to-pyrimidine ratio can be indicative of ammonia content and help constrain its abundance.

     Second, the detection of ribose would provide further context on the availability of ammonia. While our models indicate that ribose is not thermodynamically prohibited under $NH_3$-rich conditions, kinetic barriers and degradation processes may significantly limit actual abundance. Additionally, we caution that verifying the prediction may be beyond the documented capabilities of DraMS. The identification of extraterrestrial sugars in meteorites (e.g., Murchison) and returned asteroid samples (e.g., Bennu) typically relies on complex wet-chemistry protocols, including acid hydrolysis followed by manual, solution-phase derivatization procedures (e.g., Furukawa et al. 2019, 2025; Meinert et al. 2025). Barnes et al. (2021) specifically mentions detecting sugars as a science objective on Titan and other studies include sugars as molecules of interest (Boulesteix et al. 2024b); however, to our knowledge, there are no specific documented procedures for DraMS to perform such analysis. Laser desorption techniques such as matrix-assisted laser desorption/ionization (MALDI) have historically been effective for ionizing carbohydrates (Harvey 1999, 2003). If the Titan surface material (e.g., water ice; Grubisic et al. 2021) acts as a protective matrix, DraMS may be able to detect the ribose ion in LDMS mode. Chiral analysis with GC-MS introduces degradation risks associated with the pre-treatments of sugars. Therefore, definitive structural identification and chirality analysis of ribose (and other sugars) remains an open question.

     Third, the distribution of fatty acids would solidify the case for the presence of ammonia and provide context about their formation processes. In particular, detection of any fatty acids beyond butanoic acid ($C_4$) would strongly imply $NH_3$-rich conditions, whereas detection limited to butanoic acid alone would be indicative of an $NH_3$-free system. Fatty acids are ideal targets for GC-MS, with DMF-DMA derivatization effectively converting the carboxyl group (–COOH) of fatty acids into volatile methyl esters, which can be separated on the instrument's columns (Boulesteix et al. 2024b).

Fourth, beyond the individual detections, DraMS can utilize cross-class correlations to construct a more comprehensive geochemical landscape of Selk crater. Because DMF-DMA targets both carboxylic acids and amino acids, a single analysis can provide a cross-referenced validation of the ammonia availability: the detection of the unique butanoic acid spike should theoretically co-occur with a restricted amino acid profile (alanine, β-alanine, proline only; Madan & Pearce 2025) if no ammonia was present during the melt pool lifetime.

Fifth, the cytosine-to-uracil (C:U) ratio detected via LDMS could function as a test for the melt pool's duration, where a high C:U ratio implies a rapid freeze-out that preserved the thermodynamic state against kinetic deamination of cytosine to uracil.

While primarily tailored for Dragonfly's exploration of Selk crater, the methodology established in our work provides a scalable predictive framework for broader ocean worlds exploration. With appropriate refinements to initial inventories and expected reactions, this approach can support mission concepts such as POSEIDON, which targets Titan's high-latitude lakes (Rodriguez et al. 2022). Such predictive models are also helpful for strategic priorities that emphasize the need for quantitative chemical baselines to characterize impact-generated (and otherwise) aqueous niches on Titan and Enceladus (Tobie et al. 2014; Sulaiman et al. 2022).

*4.5. Strategies for Dragonfly to Seek Candidate Biosignatures*

The thermodynamic favorability of nucleobases, ribose, fatty acids, and amino acids (Madan & Pearce 2025) suggests that Selk crater may host a rich organic inventory and simultaneously implies that the mere detection of these will be consistent with abiotic chemistry. As such, the search for candidate biosignatures by Dragonfly must focus on patterns, particularly by identifying deviations from the thermodynamic and kinetic baselines established by our models and other experimental work.

Abiotic synthesis governed by chemical thermodynamics and kinetics typically produces smooth, log-normal distributions of a molecule class (e.g. fatty acids) wherein smaller and simpler molecules are more abundant than larger ones (Dorn et al. 2011). In contrast, biological systems selectively synthesize specific molecules required for function, resulting in increased abundances that may defy thermodynamic probability (Yoffe et al. 2025). For instance, our model predicts a broad, smooth distribution of fatty acids (Figure 2, Table 3), although precise smoothness may reflect the idealized equilibrium treatment of our models. A detection of this full suite with monotonically decreasing abundance would therefore suggest an abiotic origin. An additional and more stringent diagnostic lies in the structure of the saturated fatty acid chain-length distribution itself. Abiotic synthesis pathways generally yield homologous series, whereas biological lipid synthesis commonly produces pronounced even-carbon-number dominance, reflecting stepwise chain elongation mechanisms involving two-carbon addition chemistry; e.g., acetyl-CoA in terrestrial biology (e.g., Bloch 1948; Ohlrogge & Browse 1995; Gupta & Gupta 2021). Such a bias represents a higher-order biological pattern that transcends simple abundance thresholds. Therefore, a fatty acid inventory exhibiting strong even-carbon chain enrichment would constitute a compelling deviation from the abiotic baseline. Fatty acids represent a particularly high-value target for such a distributional analysis. Our results show that short-chain acids are thermodynamically accessible; however, they are insufficient for forming stable vesicle membranes which typically require chain lengths of eight carbons ($C_8$) or greater to form semi-permeable barriers in aqueous environments (Budin et al. 2014; Lai et al. 2019; Malaska et al. 2025). Therefore, the specific detection of an inventory skewed toward these longer carbon chains would also be compelling. Furthermore, we must apply lessons from recent Martian findings. Freissinet et al. (2025) reported the detection of $C_{10}$–$C_{12}$ alkanes in Gale crater mudstones, interpreting them as potential products of fatty acids. On Titan, distinguishing between alkanes and fatty acid derivatives will be crucial. This requires DraMS to characterize the entire range of chain lengths, not just the longer ones, to establish the local geological context and abiotic baseline.

Homochirality is another avenue for seeking candidate biosignatures. While our thermodynamic model does not distinguish between enantiomers, abiotic synthesis typically yields racemic mixtures. The detection of a strong enantiomeric excess (e.g., L-amino acids or D-ribose) would be a compelling signature of biological processes, though abiotic mechanisms such as circularly polarized light can induce enantiomeric excess as well (e.g., Modica et al. 2014; Glavin et al. 2020). However, part of this search may be operationally complex as the detection of ribose remains uncertain (Section 4.4).

Titan serves as a test bed for the limits of abiotic chemistry. If a world as chemically distinct as Titan can synthesize biologically relevant molecules (for terrestrial life) purely through abiotic synthesis, it compels us to re-evaluate our definitions of biosignatures. Rather than searching solely for Earth-like life or its signatures, Dragonfly's mission is equally valuable for defining the upper bound of chemical complexity achievable without biology.

*4.6. Future Work*

While our work establishes the thermodynamic feasibility of complex prebiotic chemistry at Selk crater, we do not prescribe specific reaction pathways, rates, or extents of preservation. Previous work indicates that degradation of amino acids trapped in ices (specifically glycine, alanine, and phenylalanine) by galactic cosmic ray (GCR) deposition—the dominant surface energy source on Titan—is insignificant over geological timescales (Madan et al. 2021). However, whether the broader suite of synthesized molecules persists intact, degrades, or undergoes further transformation during melt freeze or potential freeze–thaw cycling remains an open question. Future work may therefore prioritize kinetic investigations, freeze–thaw experiments, and extensions of thermodynamic models to address any of the limitations outlined in Section 2.3. Applying this modeling approach to other ocean worlds, such as Enceladus or Europa, could also allow for comparative assessments of abiotic chemical potential across diverse extraterrestrial environments.

**5. Conclusion**

Prebiotic Chemistry in Selk Crater on Titan

We present a thermodynamic assessment of nucleobases, ribose, and fatty acids within Selk's impact melt pool on Titan, completing the set of canonical molecular building blocks when including previously examined amino acids (Madan & Pearce 2025). Our key takeaways can be summarized as follows:

- Investigated molecules from all major classes of terrestrial biomolecules (amino acids, nucleobases, ribose (sugar), and fatty acids) are thermodynamically accessible in the modeled Selk melt pool environment, once ammonia is introduced (≥1 % $NH_3$ relative to water). In $NH_3$-free systems, only alanine, β-alanine, proline, adenine, and butanoic acid are favorable.
- Thermodynamic equilibrium in the models is governed by three coupled factors: stoichiometry, initial organic inventory, and Gibbs energetics. The absence of ribose, most nucleobases, and nearly all fatty acids in $NH_3$-free scenarios arises from hydrogen deficits imposed by the starting inventory. Ammonia functions as a hydrogen reservoir that enables elemental balance while minimizing Gibbs free energy.
- Distinct molecular classes exhibit different $NH_3$ optima: nucleobases, ribose, and $C_2$–$C_6$ fatty acids peak at 1% $NH_3$, whereas longer-chain fatty acids ($C_7$–$C_{12}$) require higher ammonia abundances (2%) to maximize yields.
- The modeled preference for pyrimidines over purines under $NH_3$-rich conditions, the suppression of ribose in $NH_3$-poor systems, and the monotonic decline of fatty acid yields with increasing chain length qualitatively mirror patterns observed in carbonaceous meteorites and returned asteroid samples.

From a mission perspective, the following testable predictions emerge for Dragonfly investigations at Selk crater:

- Ammonia presence and abundance may be indirectly constrained through molecular distributions:
  - The relative abundance of purines to pyrimidines provides an indicator of $NH_3$-rich vs. $NH_3$-poor aqueous processing, with $NH_3$-rich systems favoring pyrimidines over purines.
  - Detection of only butanoic acid would be consistent with $NH_3$-poor conditions, whereas detection of a broader suite of $C_2$–$C_{12}$ fatty acids, especially longer-chain species, would indicate $NH_3$-rich conditions during melt pool evolution.
  - Detection of multiple molecules spanning amino acids, nucleobases, sugars, and fatty acids would be consistent with $NH_3$-rich scenarios. In contrast, detection limited to alanine, β-alanine, proline, adenine, or butanoic acid would indicate an $NH_3$-poor environment.

On Titan, the availability of simple atmospheric organics appears sufficient to regulate thermodynamic access to the canonical building blocks of life. By establishing the equilibrium predictions for Selk's melt pool chemistry, we provide a baseline that can be used to test specific hypotheses. Molecular patterns may act as environmental diagnostics, allowing us to distinguish what Titan can achieve with abiotic chemistry from what might require alternative explanations.

Prebiotic Chemistry in Selk Crater on Titan

Defining such limits is essential as we search for life's ingredients beyond Earth and attempt to separate chemical possibility from any genuine anomaly. In doing so, Dragonfly becomes not only a search for biosignatures, but a probe of the boundaries of abiotic complexity.

**Acknowledgements**
Our research relied on computational resources provided by the Negishi cluster which is operated by the Rosen Center for Advanced Computing at Purdue University. We acknowledge the Purdue Community Cluster Program as described in (McCartney et al. 2014). We thank our anonymous reviewer and Dr. Josh Hedgepeth (second reviewer) for their thoughtful and constructive feedback that significantly improved the manuscript.

**Author Contributions**
I. Madan: Conceptualization, Data Curation, Formal Analysis, Investigation, Methodology, Project Administration, Resources, Software, Validation, Visualization, Writing – original draft, Writing – review & editing. B. Pearce: Conceptualization, Methodology, Resources, Supervision, Validation, Writing – review & editing.

**Conflicts of Interest**
None.

**Data Availability**
The necessary code(s) to reproduce any of the results in this work are openly available on Zenodo at doi:10.5281/zenodo.17445312.

**Financial Support**
I.M. was supported by the Frederick N. Andrews Fellowship awarded by Purdue University (2024-26).

**Appendix A: Gibbs Free Energies for Xanthine and Hypoxanthine**

Thermodynamic data for xanthine and hypoxanthine were not available in the CHNOSZ database; therefore, their $\Delta_f G^o$ were estimated using the Gibbs free energy estimator described in Section 2.2 of Madan and Pearce (2025), with the resulting values reported in Table A1.

Table A1. Calculated Standard Gibbs Free Energies of Formation, $\Delta_f G^o$ (J mol$^{-1}$) for Xanthine and Hypoxanthine as a function of temperature.

| | $\Delta_f G^o$ (J mol$^{-1}$) | |
|---|---|---|
| **T (ºC)** | **Xanthine** | **Hypoxanthine** |
| 0.01 | -104732 | 121455 |

Prebiotic Chemistry in Selk Crater on Titan

| | | |
|---|---|---|
| 10 | -108079 | 117898 |
| 20 | -111429 | 114338 |
| 30 | -114779 | 110778 |
| 40 | -118129 | 107218 |
| 50 | -121479 | 103658 |
| 60 | -124829 | 100098 |
| 70 | -128179 | 96538 |
| 80 | -131529 | 92978 |
| 90 | -134879 | 89418 |
| 100 | -138229 | 85858 |
| 110 | -141579 | 82298 |
| 120 | -144929 | 78738 |
| 130 | -148279 | 75178 |
| 140 | -151629 | 71618 |
| 150 | -154979 | 68058 |
| 160 | -158329 | 64498 |
| 170 | -161679 | 60938 |
| 180 | -165029 | 57378 |
| 190 | -168379 | 53818 |
| 200 | -171729 | 50258 |
| 210 | -175079 | 46698 |
| 220 | -178429 | 43138 |
| 230 | -181779 | 39578 |
| 240 | -185129 | 36018 |
| 250 | -188479 | 32458 |
| 260 | -191829 | 28898 |
| 270 | -195179 | 25338 |
| 280 | -198529 | 21778 |
| 290 | -201879 | 18218 |
| 300 | -205229 | 14658 |
| 310 | -208579 | 11098 |
| 320 | -211929 | 7538 |
| 330 | -215279 | 3978 |
| 340 | -218629 | 418 |
| 350 | -221979 | -3142 |
| 360 | -225329 | -6702 |
| 370 | -228679 | -10262 |

**Appendix B: Sensitivity Analysis**

We evaluate the sensitivity of our thermochemical model to the initial inventory of delivered organics by varying the survival factor of atmospheric deposition from the fiducial 10% to conservative (1%) (Figure B1 and Table B1) and optimistic values (30%) (Figure B1 and Table B2). This variation modifies the initial concentrations of HCN and $C_2H_2$ relative to the melt water volume.

The thermodynamic favorability of forming the studied species is preserved across all scenarios: no new species appear, nor do any fiducial species disappear. However, the magnitude of production shifts depending on molecule class. We observe an inverse relationship between survival factor and yield of nucleobases. Increasing the initial organic inventory from 1% to 30% results in a slight decrease in percent yield. Ribose, on the other hand, shows a positive correlation with reactant availability. Fatty acid behavior depends on chain length. Short fatty acids ($C_2$–$C_4$) preserve the qualitative trends of the fiducial model across all scenarios, with peak yields occurring at low $NH_3$ concentrations (≤2%). However, longer fatty acids ($C_5$–$C_{12}$) exhibit a distinct behavioral shift in the 30% survival scenario. In the 1% and 10% models, these species reach peak yields at low ammonia concentrations (~1–2%). In the 30% scenario, the production of long fatty acids is strongly suppressed at ≤2% $NH_3$ and instead peaks at higher concentrations (~3–5% $NH_3$), with the optimal $NH_3$ abundance increasing with chain length. In our model, the formation of longer hydrocarbon chains is thermodynamically disfavored relative to shorter-chain counterparts until sufficient ammonia is supplied.

Figure B1. Heat maps of percent yields (applicable to aqueous environments spanning 0–100 ºC) relative to the starting limiting reagent for (A) nucleobases, (B) ribose, and (C) $C_2$–$C_{12}$ fatty acids with 1% and 30% survival factors for the initial organic inventory. Columns correspond to varying $NH_3$ concentrations (relative to total water content). Rows list the individual molecules in each class. Cell shading follows a common viridis color scale from 0% (purple) to >70% (yellow) yield.

Prebiotic Chemistry in Selk Crater on Titan

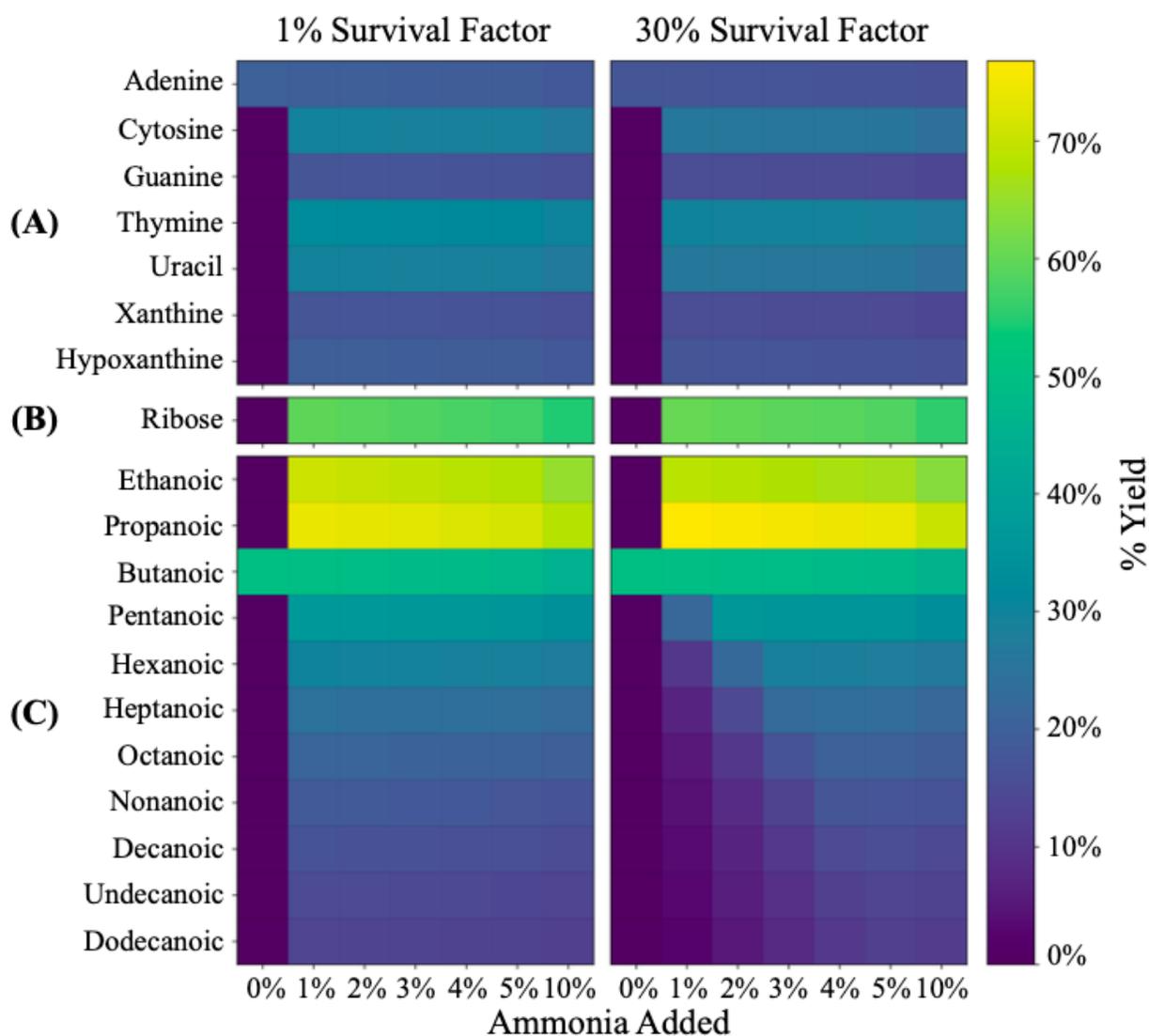

Table B1. Percent yields of nucleobases, ribose, and $C_2$–$C_{12}$ fatty acids, expressed relative to the limiting reagent (HCN for nucleobases and ribose; $C_2H_2$ for fatty acids). The survival factor is 1%.

| Molecule Class | Molecule | Yield at 1% Survival Factor | | | | | | |
|---|---|---|---|---|---|---|---|---|
| | | 0% $NH_3$ | 1% $NH_3$ | 2% $NH_3$ | 3% $NH_3$ | 4% $NH_3$ | 5% $NH_3$ | 10% $NH_3$ |
| Nucleobase | Adenine | 19.9 | 19.7 | 19.5 | 19.3 | 19.1 | 19.0 | 18.1 |
| | Cytosine | 0.0 | 29.6 | 29.3 | 29.0 | 28.7 | 28.5 | 27.2 |
| | Guanine | 0.0 | 17.4 | 17.2 | 17.1 | 16.9 | 16.7 | 16.0 |
| | Thymine | 0.0 | 32.9 | 32.6 | 32.3 | 31.9 | 31.6 | 30.2 |

Prebiotic Chemistry in Selk Crater on Titan

| Molecule Class | Molecule | 0% NH₃ | 1% NH₃ | 2% NH₃ | 3% NH₃ | 4% NH₃ | 5% NH₃ | 10% NH₃ |
|---|---|---|---|---|---|---|---|---|
| | Uracil | 0.0 | 29.6 | 29.3 | 29.0 | 28.7 | 28.5 | 27.2 |
| | Xanthine | 0.0 | 17.4 | 17.2 | 17.1 | 16.9 | 16.7 | 16.0 |
| | Hypoxanthine | 0.0 | 19.7 | 19.5 | 19.3 | 19.1 | 19.0 | 18.1 |
| **Sugar** | Ribose | 0.0 | 59.4 | 58.9 | 58.3 | 57.7 | 57.2 | 54.6 |
| | Ethanoic | 0.0 | 70.6 | 69.9 | 69.3 | 68.6 | 67.9 | 64.9 |
| | Propanoic | 0.0 | 74.3 | 73.6 | 72.9 | 72.2 | 71.5 | 68.3 |
| | Butanoic | 50.0 | 49.5 | 49.0 | 48.5 | 48.1 | 47.6 | 45.5 |
| | Pentanoic | 0.0 | 37.1 | 36.7 | 36.4 | 36.0 | 35.7 | 34.1 |
| | Hexanoic | 0.0 | 29.7 | 29.4 | 29.1 | 28.8 | 28.6 | 27.3 |
| **Fatty Acid** | Heptanoic | 0.0 | 24.7 | 24.5 | 24.2 | 24.0 | 23.8 | 22.7 |
| | Octanoic | 0.0 | 21.2 | 21.0 | 20.8 | 20.6 | 20.4 | 19.5 |
| | Nonanoic | 0.0 | 18.5 | 18.4 | 18.2 | 18.0 | 17.8 | 17.0 |
| | Decanoic | 0.0 | 16.5 | 16.3 | 16.2 | 16.0 | 15.9 | 15.1 |
| | Undecanoic | 0.0 | 14.8 | 14.7 | 14.5 | 14.4 | 14.3 | 13.6 |
| | Dodecanoic | 0.0 | 13.5 | 13.4 | 13.2 | 13.1 | 13.0 | 12.4 |

Table B2. Percent yields of nucleobases, ribose, and $C_2$–$C_{12}$ fatty acids, expressed relative to the limiting reagent (HCN for nucleobases and ribose; $C_2H_2$ for fatty acids). The survival factor is 30%.

| Molecule Class | Molecule | Yield at 30% Survival Factor | | | | | | |
|---|---|---|---|---|---|---|---|---|
| | | 0% NH₃ | 1% NH₃ | 2% NH₃ | 3% NH₃ | 4% NH₃ | 5% NH₃ | 10% NH₃ |
| | Adenine | 17.6 | 17.4 | 17.3 | 17.1 | 17 | 16.8 | 16.2 |
| | Cytosine | 0 | 26.4 | 26.2 | 26.0 | 25.8 | 25.5 | 24.5 |
| **Nucleobase** | Guanine | 0 | 15.4 | 15.3 | 15.1 | 15.0 | 14.9 | 14.3 |
| | Thymine | 0 | 29.9 | 29.6 | 29.4 | 29.1 | 28.9 | 27.7 |
| | Uracil | 0 | 26.4 | 26.2 | 26.0 | 25.8 | 25.5 | 24.5 |
| | Xanthine | 0 | 15.4 | 15.3 | 15.1 | 15.0 | 14.9 | 14.3 |

Prebiotic Chemistry in Selk Crater on Titan

|  |  |  |  |  |  |  |  |  |
|---|---|---|---|---|---|---|---|---|
|  | Hypoxanthine | 0 | 17.4 | 17.3 | 17.1 | 17.0 | 16.8 | 16.2 |
| **Sugar** | Ribose | 0 | 60.5 | 59.9 | 59.3 | 58.8 | 58.2 | 55.5 |
| **Fatty Acid** | Ethanoic | 0 | 68.9 | 68.2 | 67.6 | 67.0 | 66.4 | 63.4 |
|  | Propanoic | 0 | 76.8 | 76.0 | 75.3 | 74.5 | 73.8 | 70.3 |
|  | Butanoic | 50.1 | 49.6 | 49.1 | 48.7 | 48.2 | 47.7 | 45.6 |
|  | Pentanoic | 0 | 22.1 | 36.3 | 35.9 | 35.6 | 35.3 | 33.7 |
|  | Hexanoic | 0 | 10.6 | 22.3 | 28.5 | 28.2 | 28.0 | 26.7 |
|  | Heptanoic | 0 | 7.0 | 14.5 | 22.5 | 23.4 | 23.2 | 22.1 |
|  | Octanoic | 0 | 5.2 | 10.7 | 16.6 | 20.0 | 19.8 | 18.9 |
|  | Nonanoic | 0 | 4.1 | 8.5 | 13.1 | 17.4 | 17.3 | 16.5 |
|  | Decanoic | 0 | 3.4 | 7.0 | 10.8 | 14.7 | 15.3 | 14.6 |
|  | Undecanoic | 0 | 3.0 | 6.0 | 9.2 | 12.5 | 13.7 | 13.1 |
|  | Dodecanoic | 0 | 2.6 | 5.3 | 8.0 | 10.9 | 12.5 | 11.9 |

**References**

Artemieva, N., & Lunine, J. 2003, Icarus, 164, 471, doi:10.1016/s0019-1035(03)00148-9
Barnes, J. W., Turtle, E. P., Trainer, M. G., et al. 2021, Planet Sci J, 2, 130, doi:10.3847/psj/abfdcf
Becke, A. D. 1993, J Chem Phys, 98, 5648, doi:10.1063/1.464913
Bloch, K. 1948, Cold Spring Harb Symp Quant Biol, 13, 29, doi:10.1101/sqb.1948.013.01.008
Boulesteix, D., Buch, A., Masson, G., et al. 2024a, AGUFM, 2024, P41B, https://ui.adsabs.harvard.edu/abs/2024AGUFMP41B...06B/abstract
Boulesteix, D., Buch, A., Samson, J., et al. 2024b, J Chromatogr A, 1722, 464860, doi:10.1016/j.chroma.2024.464860
Boyer, G. 2024, *pyCHNOSZ: Python wrapper for the thermodynamic package CHNOSZ* (Zenodo)
Brassé, C., Buch, A., Coll, P., & Raulin, F. 2017, Astrobiology, 17, 8, doi:10.1089/ast.2016.1524
Budin, I., Prwyes, N., Zhang, N., & Szostak, J. W. 2014, Biophys J, 107, 1582, doi:10.1016/j.bpj.2014.07.067
Butlerow, A. 1861, Justus Liebigs Ann Chem, 120, 295, https://scholar.archive.org/work/qgxmsb2cvvgmjcadffzhnve4gi/access/ia_file/crossref-pre-1909-scholarly-works/10.1002%252Fjlac.18601150302.zip/10.1002%252Fjlac.18611200308.pdf
Cable, M. L., Hörst, S. M., Hodyss, R., et al. 2012, Chem Rev, 112, 1882, doi:10.1021/cr200221x
Cammi, R., & Tomasi, J. 1995, J Comput Chem, 16, 1449, doi:10.1002/jcc.540161202
Cappelletti, M., & Rahm, M. 2026, J Am Chem Soc, doi:10.1021/jacs.5c09522
Choukroun, M., & Sotin, C. 2012, Geophys Res Lett, 39, doi:10.1029/2011gl050747
Cleaves, H. J., II, Neish, C., Callahan, M. P., et al. 2014, Icarus, 237, 182, doi:10.1016/j.icarus.2014.04.042
Cobb, A. K., & Pudritz, R. E. 2014, Astrophys J, 783, 140, doi:10.1088/0004-637x/783/2/140
Cobb, A. K., Pudritz, R. E., & Pearce, B. K. D. 2015, Astrophys J, 809, 6, doi:10.1088/0004-

Prebiotic Chemistry in Selk Crater on Titan

637x/809/1/6
Committee on the Planetary Science and Astrobiology Decadal Survey, Space Studies Board, Division on Engineering and Physical Sciences, & National Academies of Sciences, Engineering, and Medicine 2023, Origins, worlds, and life, doi:10.17226/27209
Cottini, V., Nixon, C. A., Jennings, D. E., et al. 2012, Planet Space Sci, 60, 62, doi:10.1016/j.pss.2011.03.015
Coustenis, A. 1997, Adv Space Res, 19, 1288, doi:10.1016/s0273-1177(97)83127-4
Cronin, J. R., Cooper, G. W., & Pizzarello, S. 1995, Adv Space Res, 15, 91, doi:10.1016/s0273-1177(99)80068-4
Crósta, A. P., Silber, E. A., Lopes, R. M. C., et al. 2021, Icarus, 370, 114679, doi:10.1016/j.icarus.2021.114679
Cynn, H. C., Boone, S., Koumvakalis, A., Nicol, M., & Stevenson, D. J. 1989, (adsabs.harvard.edu)
Dick, J. M. 2019, Front Earth Sci, 7, doi:10.3389/feart.2019.00180
Dorn, E. D., Nealson, K. H., & Adami, C. 2011, J Mol Evol, 72, 283, doi:10.1007/s00239-011-9429-4
Engel, M. H., & Nagy, B. 1982, Nature, 296, 837, doi:10.1038/296837a0
Engel, S., Lunine, J. I., & Norton, D. L. 1994, J Geophys Res, 99, 3745, doi:10.1029/93je03433
Farnsworth, K. K., McLain, H. L., Chung, A., & Trainer, M. G. 2024, ACS Earth Space Chem, 8, 2380, doi:10.1021/acsearthspacechem.4c00114
Ferris, J. P., Joshi, P. C., Edelson, E. H., & Lawless, J. G. 1978, J Mol Evol, 11, 293, doi:10.1007/bf01733839
Freissinet, C., Glavin, D. P., Archer, P. D., Jr, et al. 2025, Proc Natl Acad Sci U S A, 122, e2420580122, doi:10.1073/pnas.2420580122
Freissinet, C., Moulay, V., Li, X., et al. 2024, ACS Earth Space Chem, doi:10.1021/acsearthspacechem.4c00143
Frisch, M. J., Trucks, G. W., Schlegel, H. B., et al. 2017, Gaussian 16, Revision B.01
Fulchignoni, M., Ferri, F., Angrilli, F., et al. 2005, Nature, 438, 785, doi:10.1038/nature04314
Furukawa, Y., Chikaraishi, Y., Ohkouchi, N., et al. 2019, Proc Natl Acad Sci U S A, 116, 24440, doi:10.1073/pnas.1907169116
Furukawa, Y., Sunami, S., Takano, Y., et al. 2025, Nat Geosci, 1, doi:10.1038/s41561-025-01838-6
Glavin, D. P., & Bada, J. L. 2001, Astrobiology, 1, 259, doi:10.1089/15311070152757456
Glavin, D. P., Bada, J. L., Brinton, K. L., & McDonald, G. D. 1999, Proc Natl Acad Sci U S A, 96, 8835, doi:10.1073/pnas.96.16.8835
Glavin, D. P., Burton, A. S., Elsila, J. E., Aponte, J. C., & Dworkin, J. P. 2020, Chem Rev, 120, 4660, doi:10.1021/acs.chemrev.9b00474
Glavin, D. P., Dworkin, J. P., Alexander, C. M. O., et al. 2025, Nat Astron, 9, 199, doi:10.1038/s41550-024-02472-9
Goodwin, D. G., Moffat, H. K., Schoegl, I., Speth, R. L., & Weber, B. W. 2024, *Cantera: An object-oriented software toolkit for chemical kinetics, thermodynamics, and transport processes* (Zenodo)
Griffith, C. A., Owen, T., Geballe, T. R., Rayner, J., & Rannou, P. 2003, Science, 300, 628, doi:10.1126/science.1081897
Grubisic, A., Trainer, M. G., Li, X., et al. 2021, Int J Mass Spectrom, 470, 116707, doi:10.1016/j.ijms.2021.116707
Gupta, R., & Gupta, N. 2021, in Fundamentals of Bacterial Physiology and Metabolism (Singapore: Springer Singapore), 491
Harvey, D. J. 1999, Mass Spectrom Rev, 18, 349, doi:10.1002/(SICI)1098-2787(1999)18:6%3C349::AID-MAS1%3E3.0.CO;2-H
Harvey, D. J. 2003, Int J Mass Spectrom, 226, 1, doi:10.1016/s1387-3806(02)00968-5
He, Y., Buch, A., Morisson, M., et al. 2019, Talanta, 204, 802, doi:10.1016/j.talanta.2019.06.076
Hedgepeth, J. E., Buffo, J. J., Chivers, C. J., Neish, C. D., & Schmidt, B. E. 2022, Planet Sci J, 3, 51, doi:10.3847/psj/ac4d9c
Hedgepeth, J. E., Neish, C. D., Turtle, E. P., et al. 2020, Icarus, 344, 113664,

Prebiotic Chemistry in Selk Crater on Titan

doi:10.1016/j.icarus.2020.113664
Hörst, S. M. 2017, J Geophys Res Planets, 122, 432, doi:10.1002/2016je005240
Janssen, M. A., Le Gall, A., Lopes, R. M., et al. 2016, Icarus, 270, 443, doi:10.1016/j.icarus.2015.09.027
Jennings, D. E., Tokano, T., Cottini, V., et al. 2019, Astrophys J Lett, 877, L8, doi:10.3847/2041-8213/ab1f91
Kalousová, K., & Sotin, C. 2020, Geophys Res Lett, 47, doi:10.1029/2020gl087481
Kalousová, K., Wakita, S., Sotin, C., et al. 2024, J Geophys Res Planets, 129, doi:10.1029/2023je008107
Khare, B., Sagan, C., Ogino, H., et al. 1986, Icarus, 68, 176, doi:10.1016/0019-1035(86)90080-1
Koga, T., Oba, Y., Takano, Y., et al. 2025, A complete set of canonical nucleobases in asteroid Ryugu demonstrates their prevalence in carbonaceous asteroids, Research Square, doi:10.21203/rs.3.rs-7071403/v1
Krasnopolsky, V. A. 2009, Icarus, 201, 226, doi:10.1016/j.icarus.2008.12.038
Lai, J. C.-Y., Pearce, B. K. D., Pudritz, R. E., & Lee, D. 2019, Icarus, 319, 685, doi:10.1016/j.icarus.2018.09.028
Laize-Générat, L., Soussaintjean, L., Poch, O., et al. 2024, Geochim Cosmochim Acta, 387, 111, doi:10.1016/j.gca.2024.10.001
LaRowe, D. E., & Regnier, P. 2008, Orig Life Evol Biosph, 38, 383, doi:10.1007/s11084-008-9137-2
Lee, C., Yang, W., & Parr, R. G. 1988, Phys Rev B Condens Matter, 37, 785, doi:10.1103/physrevb.37.785
Leitner, M. A., & Lunine, J. I. 2019, Icarus, 333, 61, doi:10.1016/j.icarus.2019.05.008
Lopes, R. M. C., Elachi, C., Mueller-Wodarg, I., & Solomonidou, A. 2025, *Titan after cassini-Huygens* (Philadelphia, PA: Elsevier - Health Sciences Division)
Lora, J. M., Turtle, E. P., & Mitchell, J. L. 2025, in Titan After Cassini-Huygens (Elsevier), 201
Lorenz, R. D., MacKenzie, S. M., Neish, C. D., et al. 2021, Planet Sci J, 2, 24, doi:10.3847/psj/abd08f
Lunine, J. I., & Stevenson, D. J. 1987, Icarus, 70, 61, doi:10.1016/0019-1035(87)90075-3
Lunine, J., Tobie, G., Horst, S., & Mandt, K. 2025, Titan After Cassini-Huygens, https://www.sciencedirect.com/science/article/pii/B9780323991612000085
Madan, I. 2025, Dataset for "Prebiotic Chemistry Insights for Dragonfly: Thermodynamics of amino acid synthesis in Selk crater on Titan," doi:10.5281/ZENODO.17445312
Madan, I., Aliabadi, S. A., Huhtasaari, J., et al. 2025, QRB Discov, 6, e23, doi:10.1017/qrd.2025.10012
Madan, I., Collins, G., & Cable, M. 2021, (Authorea)
Madan, I., & Pearce, B. K. D. 2025, Planet Sci J, 6, 284, doi:10.3847/psj/ae1c18
Malaska, M. J., Sandström, H., Hofmann, A. E., et al. 2025, Astrobiology, 25, 367, doi:10.1089/ast.2024.0125
Martins, Z., & Sephton, M. A. 2009, *Amino acids, peptides and proteins in organic chemistry: Origins and synthesis of amino acids* (1st ed.; Weinheim, Germany: Wiley-VCH Verlag)
Masson, G., Buch, A., Boulesteix, D., et al. 2025, (Copernicus Meetings)
McBride, B. J., Zehe, M. J., & Gordon, S. 2002, http://200.17.228.88/CFD/bibliografia/propulsao/NASA_TP-2002-211556.pdf
McCartney, G., Hacker, T., & Yang, B. 2014, Empowering Faculty: A Campus Cyberinfrastructure Strategy for Research communities, EDUCAUSE review
Meinert, C., Leyva, V., Robert, M., Pepino, R., & Bocková, J. 2025, Abiotic sugars in (162173) Ryugu and the primitive CI carbonaceous chondrite Orgueil, Research Square, doi:10.21203/rs.3.rs-6916147/v1
Meyer-Dombard, D. R., Malas, J., Russo, D. C., & Kenig, F. 2025, in Titan After Cassini-Huygens, ed. R. M. C. Lopes, C. Elachi, I. C. F. Müeller-Wodarg, & A. Solomonidou (Elsevier), 423
Miertuš, S., Scrocco, E., & Tomasi, J. 1981, Chem Phys, 55, 117, doi:10.1016/0301-0104(81)85090-2
Missen, R. W., & Smith, W. R. 1998, *Chemical Reaction Stoichiometry (CRS): A Tutorial*
Modica, P., Meinert, C., de Marcellus, P., et al. 2014, The Astrophysical Journal, 788, 79, doi:10.1088/0004-637X/788/1/79
Mojarro, A., Aponte, J. C., Dworkin, J. P., et al. 2025, Proc Natl Acad Sci U S A, 122, e2512461122, doi:10.1073/pnas.2512461122

Prebiotic Chemistry in Selk Crater on Titan

Naraoka, H., Takano, Y., Dworkin, J. P., et al. 2023, Science, 379, eabn9033, doi:10.1126/science.abn9033
Nasralla, M., Laurent, H., Alderman, O. L. G., & Dougan, L. 2025, Commun Chem, 8, 227, doi:10.1038/s42004-025-01599-8
Neish, C. D., Barnes, J. W., Sotin, C., et al. 2015, Geophys Res Lett, 42, 3746, doi:10.1002/2015GL063824
Neish, C. D., & Lorenz, R. D. 2012, Planet Space Sci, 60, 26, doi:10.1016/j.pss.2011.02.016
Neish, C. D., Lorenz, R. D., Turtle, E. P., et al. 2018, Astrobiology, 18, 571, doi:10.1089/ast.2017.1758
Neish, C. D., Molaro, J. L., Lora, J. M., et al. 2016, Icarus, 270, 114, doi:10.1016/j.icarus.2015.07.022
Neish, C. D., Somogyi, Á., Lunine, J. I., & Smith, M. A. 2009, Icarus, 201, 412, doi:10.1016/j.icarus.2009.01.003
Neish, C. D., Somogyi, A., & Smith, M. A. 2010, Astrobiology, 10, 337, doi:10.1089/ast.2009.0402
Neish, C., Malaska, M. J., Sotin, C., et al. 2024, Astrobiology, 24, 177, doi:10.1089/ast.2023.0055
Niemann, H., Atreya, S., Bauer, S., et al. 2005, Nature, 438, 779, doi:10.1038/nature04122
Niemann, H. B., Atreya, S. K., Demick, J. E., et al. 2010, Journal of Geophysical Research: Planets, 115, doi:10.1029/2010JE003659
Oba, Y., Koga, T., Takano, Y., et al. 2023, Nat Commun, 14, 1292, doi:10.1038/s41467-023-36904-3
Oba, Y., Takano, Y., Furukawa, Y., et al. 2022, Nat Commun, 13, 2008, doi:10.1038/s41467-022-29612-x
Ochterski, J. W. 2000, Gaussian Inc, 1, https://www.researchgate.net/profile/Prem-Baboo/post/How-to-create-catalyst-surface-in-gaussian-to-generate-input-file-along-with-the-reactants/attachment/59d634ad79197b807799254c/AS%3A380803725971456%401467802091328/download/G98thermo.pdf
Ohlrogge, J., & Browse, J. 1995, Plant Cell, 7, 957, doi:10.1105/tpc.7.7.957
Oro, J., & Kimball, A. P. 1961, Arch Biochem Biophys, 94, 217, doi:10.1016/0003-9861(61)90033-9
Parker, E. T., McLain, H. L., Glavin, D. P., et al. 2025, in 56th Lunar and Planetary Science Conference (LPSC) (ntrs.nasa.gov)
Paschek, K., Kohler, K., Pearce, B. K. D., et al. 2022, Life (Basel), 12, 404, doi:10.3390/life12030404
Pearce, B. K. D., Hörst, S. M., Cline, C. J., et al. 2024, Planet Sci J, 5, 68, doi:10.3847/psj/ad283e
Pearce, B. K. D., & Pudritz, R. E. 2016, Astrobiology, 16, 853, doi:10.1089/ast.2015.1451
Perrin, Z., Carrasco, N., Gautier, T., et al. 2025, Icarus, 429, 116418, doi:10.1016/j.icarus.2024.116418
Petricca, F., Vance, S. D., Parisi, M., et al. 2025, Nature, 648, 556, doi:10.1038/s41586-025-09818-x
Poch, O., Coll, P., Buch, A., Ramírez, S. I., & Raulin, F. 2012, Planet Space Sci, 61, 114, doi:10.1016/j.pss.2011.04.009
Potiszil, C., Ota, T., Yamanaka, M., et al. 2023a, Nat Commun, 14, 1482, doi:10.1038/s41467-023-37107-6
Potiszil, C., Yamanaka, M., Sakaguchi, C., et al. 2023b, Life (Basel), 13, 1448, doi:10.3390/life13071448
Raulin, F., & Owen, T. 2003, in The Cassini-Huygens Mission (Dordrecht: Springer Netherlands), 377
Rodriguez, S., Vinatier, S., Cordier, D., et al. 2022, Exp Astron, 54, 911, doi:10.1007/s10686-021-09815-8
Sephton, M. A. 2002, Nat Prod Rep, 19, 292, doi:10.1039/b103775g
Sillerud, L. O. 2024, in Abiogenesis (Cham: Springer Nature Switzerland), 781
Solomonidou, A., Coustenis, A., Gall, A. L., et al. 2024, in IGARSS 2024 - 2024 IEEE International Geoscience and Remote Sensing Symposium (IEEE), 6092
Solomonidou, A., Le Gall, A., Hayne, P., & Coustenis, A. 2025, in Titan After Cassini-Huygens, ed. R. M. C. Lopes, C. Elachi, I. C. F. Müeller-Wodarg, & A. Solomonidou (Elsevier), 325
Solomonidou, A., Neish, C., Coustenis, A., et al. 2020, Astron Astrophys, 641, A16, doi:10.1051/0004-6361/202037866
Stern, J. C., Trainer, M. G., Brinckerhoff, W. B., et al. 2023, in 54th LPSC Lunar and Planetary Science Conference 2023 (insu.hal.science), LPI Contribution No. 2806
Sulaiman, A. H., Achilleos, N., Bertucci, C., et al. 2022, Exp Astron, 54, 849, doi:10.1007/s10686-021-

09810-z

Thompson, W. R., & Sagan, C. 1992, ESASP, 338, 167, https://ui.adsabs.harvard.edu/abs/1992ESASP.338..167T/abstract

Tobie, G., Grasset, O., Lunine, J. I., Mocquet, A., & Sotin, C. 2005, Icarus, 175, 496, doi:10.1016/j.icarus.2004.12.007

Tobie, G., Teanby, N. A., Coustenis, A., et al. 2014, Planet Space Sci, 104, 59, doi:10.1016/j.pss.2014.10.002

Tomasi, J., Mennucci, B., & Cancès, E. 1999, Journal of Molecular Structure: THEOCHEM, 464, 211, doi:10.1016/S0166-1280(98)00553-3

Vuitton, V., Lavvas, P., Nixon, C. A., & Teanby, N. A. 2025, in Titan After Cassini-Huygens, ed. R. M. C. Lopes, C. Elachi, I. C. F. Müeller-Wodarg, & A. Solomonidou (Elsevier), 157

Waite, J. H., Jr, Young, D. T., Cravens, T. E., et al. 2007, Science, 316, 870, doi:10.1126/science.1139727

Wakita, S., Johnson, B. C., Soderblom, J. M., & Neish, C. D. 2024, LPI Contrib, 3040, 1146, https://ui.adsabs.harvard.edu/abs/2024LPICo3040.1146W/abstract

Wakita 脇田, S. 茂., Johnson, B. C., Soderblom, J. M., et al. 2023, Planet Sci J, 4, 51, doi:10.3847/PSJ/acbe40

Werynski, A., Neish, C. D., Gall, A. L., & Janssen, M. A. 2019, Icarus, 321, 508, doi:10.1016/j.icarus.2018.12.007

Yoffe, G., Klenner, F., Sober, B., Kaspi, Y., & Halevy, I. 2025, Molecular diversity as a biosignature, arXiv [astro-ph.EP], doi:10.48550/arXiv.2511.00525